\documentclass[12pt]{article}

\input epsf
\def\beq{\begin{eqnarray}}
\def\eeq{\end{eqnarray}}

\def\leh{{\cal L}^{(2)}}
\def\L*{{\cal L}_*}
\def\lsim{\mathrel{\rlap{\lower3pt\hbox{\hskip0pt$\sim$}}
     \raise1pt\hbox{$<$}}}         
\def\gsim{\mathrel{\rlap{\lower4pt\hbox{\hskip1pt$\sim$}}
     \raise1pt\hbox{$>$}}}         

\usepackage{amsmath}
\usepackage{amsfonts}

\begin{document}

\begin{titlepage}

\begin{flushright}
{NYU-TH-07/12/30}
\end{flushright}
\vskip 0.9cm

\centerline{\Large \bf Special Massive Spin-2 on de Sitter Space}

\vskip 0.7cm
\centerline{\large Gregory Gabadadze$^a$ and Alberto Iglesias$^b$}
\vskip 0.3cm
\centerline{$^a$\em Center for Cosmology and Particle Physics}
\centerline{\em Department of Physics, New York University, New York, 
NY, 10003, USA}
\centerline{$^b$\em Department of Physics, University of California, Davis,
CA 95616, USA}

\vskip 1.9cm

\begin{abstract}

The theory of a  massive spin-2 state on the de Sitter space -- 
with the mass squared equal to one sixth of the curvature -- 
is special for two reasons: (i) it exhibits 
an enhanced local symmetry;  (ii) it emerges as a part of the 
model that gives rise to the self-accelerated Universe. 
The known problems of this theory are:  either it 
cannot be coupled to a non-conformal conserved stress-tensor 
because of the enhanced symmetry, or  it propagates a 
ghost-like state when the symmetry is constrained 
by the Lagrange multiplier method.  Here we propose a solution  to 
these problems in the linearized approximation.

\end{abstract}

\vspace{3cm}

\end{titlepage}

\newpage

\section{Introduction and summary}

The main topic of the present note is a theory of massive  
tensor perturbation on the de Sitter (dS) background, 
when there is a special relation between the mass of the 
perturbation, $m_*$, and the  curvature of the background, ${\bar R}$:
\beq
m_*^2= {{\bar R}\over 6}\,.
\label{mstar0}
\eeq
Our  interest in such a theory is twofold:

(I) It has been known since  the works of Deser and Nepomechie 
\cite {Deser}, and  Higuchi \cite {Higuchi} (see also \cite {DeserWaldron}), 
that the linearized theory of a massive graviton on the  dS  background 
with the mass-curvature relation  (\ref {mstar0}) exhibits an 
additional local symmetry. The corresponding symmetry transformations are 
generalization of conformal transformations; the symmetry 
disappears for any  other value of $m_*$.  As a consequence  of 
this symmetry, the tensor mode (graviton) 
of mass (\ref {mstar0}) propagates on the dS background 
four on-shell helicity states \cite {Deser,DeserWaldron}. 
However, such a theory cannot be coupled consistently to 
a conserved stress-tensor with a non-zero trace \cite {Higuchi}. 
The question is whether this latter drawback can be improved.

(II) In the Dvali-Gabadadze-Porrati (DGP) model of  modified gravity 
\cite {DGP}, there is a 
self-accelerated solution which  could  model the accelerated expansion 
of the  Universe 
\cite {Cedric,DDG}. The acceleration there is due to ``dark energy'' 
which is supplied by gravity itself, in particular, by a nonzero 
condensate of a massive graviton.  Interestingly enough, the (lightest) 
graviton  mass on the self-accelerated background  
is related to the curvature as in (\ref {mstar0}) just 
automatically \cite {Koyama}. 
This is because both the graviton mass and 
curvature there  have the same origin.
The model exhibits the strongly coupled behavior 
at the nonlinear level \cite {Arkady,DDGV} (see also, \cite {Rat1,Rubakov}).  
In the linearized theory the problem  of coupling to a conserved  
stress-tensor of nonzero trace, mentioned in the previous 
paragraph, is absent. However, the corresponding linearized theory  
on the self-accelerated background has a ghost-like state 
\cite {Rat1,Rat2,Gorbunov,Kaloper,KKTanaka}. 
Because the theory is in the strongly coupled 
regime \cite {DDGV,Rat1,Rubakov}, 
the results of the linearized theory  remain unwarranted  
in the full non-linear theory \cite {DGI} (see also \cite {Dvali}).
Nevertheless, the  presence of the ghost in the linearized theory is 
indicative \cite {Cedric,Rat1} of potential negative mass solutions 
in the full nonlinear theory. Such negative mass solutions have been 
obtained  in a semi-exact \cite {GI} and exact nonperturbative 
settings in the DGP model \cite {DGPR} (see also,
\cite{KaloperMyers}). This suggest that the self-accelerated 
background should be unstable, however, it is not clear 
whether the instability is rapid or not; 
for instance, an explicit calculation of the decay of the 
selfaccelerated  branch into  the conventional branch  
shows that such a  decay does not take place, at lest in a 
quasi-classical approximation \cite {Bounce}. 

The present note concerns a simpler issue. Here, 
we will consider the linearized theory 
and will try to decouple the ghost-like state from the 
physical sector. A consistent nonlinear theory
of a massive graviton on dS space with the 
mass-curvature relation  (\ref {mstar0}), if exists, 
should in the linearized approximation reduce to the 
Lagrangian given  in the present paper.
 
We will show that  it is possible to have a fully consistent linearized 
theory  of massive spin-2 on dS background with the 
mass-curvature relation (\ref {mstar0}). We show
this by explicitly setting up the Lagrangian of such a theory 
and calculating the physical one-graviton exchange 
amplitude, from which the tachyon and ghost states decouple. 
As to its nonlinear completion, the full theory may or 
may not be strongly coupled, depending on the 
value of a certain coupling\footnote{
The perturbative stability of massive gravity on
de Sitter has also been considered albeit within the
context of bi-gravity theories in \cite{Blas:2007zz}.}.

The organization of the paper is as follows:
In section 2 we summarize the results on  
the enhanced symmetry of massive graviton on dS background 
with the mass-curvature relation  (\ref {mstar0}).
Then we discuss additional terms that appear in the low-energy 
Lagrangian of the DGP model on the self-accelerated 
background \cite {Gorbunov},  this section serves to clarify our  
computational tools, which in the case of 
tensor fields on curved background are subtle. In section 3
we discuss a further extension of the linearized Lagrangian 
which allows to decouple the ghost and tachyons, and calculate the  
one-graviton exchange amplitude. We give brief comments 
on the issues associated with non-linear interactions.

\section{Enhanced symmetry {\bf \it vs.} a ghost}

Consider an expansion about 
the conventional de Sitter background of curvature 
${\bar R}=12H^2$. The perturbations of the metric 
are  defined as follows:
\beq
g_{\mu\nu} = \gamma_{\mu\nu} + h_{\mu\nu}\,,
\label{pert}
\eeq
where $\gamma_{\mu\nu}$ denotes the background dS metric.
The Lagrangian that we are interested in takes the form \cite {Higuchi}
(we set $8\pi G_N=1$, with $G_N$ being  the Newton constant)
\beq
\L* \equiv  \leh  - {1\over 4} m_*^2 \left ( h_{\mu\nu}^2 -h^2 \right )\,, 
\label{lstar}
\eeq
where $m_*$ is the graviton mass and $\leh $
denotes the Einstein-Hilbert Lagrangian expanded up to the  
quadratic order about the dS background:
\beq
\leh \equiv {1\over 2} (\nabla_\mu h^{\mu\alpha})^2  +
{1\over 4}h_{\mu\nu} \square   h^{\mu\nu} 
-{1\over 4} h \square  h +{1\over 2} h^{\mu\nu} \nabla_\mu \nabla_\nu h 
- {1\over 2} H^2 \left ( h^2_{\mu\nu} +{1\over 2} h^2  \right )\,.
\label{EH}
\eeq
All the covariant derivatives above are taken w.r.t.~the background dS 
metric,  and covariant d'Alambertian is denoted by 
$\square \equiv \gamma^{\mu\nu}\nabla_\mu \nabla_\nu$.

Because of the mass term, the Lagrangian (\ref {lstar}) is not invariant 
w.r.t.~linearized reperametrizations, $h_{\mu\nu}\to h_{\mu\nu}+
\nabla_\mu \zeta_\nu + \nabla_\nu \zeta_\mu $, with  $\zeta_\mu$ being
a transformation parameter. However, the 
symmetry can be  restored by using the St\"uckelberg  method:
redefine the field as $h_{\mu\nu} \equiv {\tilde h}_{\mu\nu} - \nabla_\mu 
V_\nu
-\nabla_\nu V_\mu $; the resulting Lagrangian written in terms of 
${\tilde h} $ and $V$ will be invariant under the simultaneous 
transformations  ${\tilde h}_{\mu\nu}\to {\tilde h}_{\mu\nu}+\nabla_\mu 
\zeta_\nu + \nabla_\nu \zeta_\mu $ and 
$V_\mu \to  V_\mu + \zeta_\mu $. Hence, we can regard 
the Lagrangian (\ref {lstar}) as a gauge fixed version of the 
reparametrization invariant theory in which the gauge condition 
$V_\mu =0$ had been enforced.

The Lagrangian (\ref {lstar}) exhibits an additional local symmetry 
\cite {Deser} when 
\beq
m_*^2 =2H^2\,.
\label{mstar}
\eeq 
The corresponding symmetry transformation is 
\beq
h_{\mu\nu}(x) \to  h_{\mu\nu}(x) +
(\nabla_\mu \nabla_\nu + H^2 \gamma_{\mu\nu}) \rho (x)\,.
\label{enhanced}
\eeq
As a consequence of this symmetry, there are four on-shell 
degrees of freedom propagated by such a graviton; moreover,  
no ghost or tachyon states appear in the spectrum  
\cite {Deser,Higuchi,DeserWaldron}.  For  $m_*<2H^2$, on the other hand, 
a ghost appears and the theory is inconsistent \cite {Higuchi}, while 
for  $m_*>2H^2$ the graviton propagates five helicity states
and the theory is unitary.  The case with the 
mass-curvature relation  (\ref {mstar}) (or with (\ref {mstar0})) is the 
main subject of this note.

So far we have been discussing pure gravity.
How about its coupling to other fields? 
The coupling can be introduced by adding to
(\ref {lstar})  the conventional term $h_{\mu\nu}T^{\mu\nu}$,
where the stress-tensor $T^{\mu\nu}$ is 
covariantly conserved, $\nabla{^\mu}T_{\mu\nu} =0$.
This does not affect our arguments about the linearized 
reparametrizations.
However, the coupling  breaks the invariance of the theory 
w.r.t.~(\ref {enhanced}), unless $T=0$. 
One could introduce a coupling {\it a la} St\"uckelberg
$(h_{\mu\nu}- (\nabla_\mu \nabla_\nu + H^2 \gamma_{\mu\nu})\sigma) 
T^{\mu\nu}$, where $\sigma$ is a field that also transforms 
as $ \delta \sigma(x) = \rho(x)$, when the metric is transformed according to 
(\ref {enhanced}). Such a coupling would not violate the symmetry of 
the theory. However, the equation of motion of the $\sigma$ field would 
require that $T^\mu_\mu=0$. 

The same problem arises if one looks at the Einstein equations following
from (\ref {lstar}): the
Bianchi identities then enforce the condition 
$( \nabla^\mu h_{\mu\nu} -\nabla_\nu h) =0$, and, on this condition,  
the trace of the Einstein equation reads 
\beq
({m_*^2} -2H^2) h = -2T\,.
\label{trace}
\eeq
Therefore, for the special value of the mass 
(\ref {mstar}), the theory does not admit 
coupling to a non-conformal source, just like the
Maxwell theory does not admit coupling to a 
non-conserved source.

How can this be changed? Our approach below is motivated by the  
low energy effective lagrangian of the DGP model on the selfaccelerated
solution which was derived in \cite {Gorbunov}.  
We consider the lagrangian of the following form (see, also \cite {DGI}):
\beq
{\cal L}_{\rm eff}={\cal L}_*(h_{\mu\nu})- \phi 
{\cal {\hat O}}^{\mu\nu}h_{\mu\nu}+h_{\mu\nu} T^{\mu\nu},
\label{eff0}
\eeq
where  we have introduced a Lagrange multiplier field $\phi$, and   
used the notation:
\beq
{\cal {\hat O}}_{\mu\nu}\equiv \nabla_\mu\nabla_\nu
-\gamma_{\mu\nu}\Box-3H^2\gamma_{\mu\nu}\,.
\label{defO}
\eeq
Irrespective of its origin, the form of $\cal {\hat O}_{\mu\nu}$ 
can be motivated by the requirement that the Bianchi identities 
be free of the field $\phi$ (see below). 

Variation w.r.t.~$\phi$ gives 
\beq
{\cal {\hat O}}^{\mu\nu}h_{\mu\nu}=0\,.
\label{Oh}
\eeq
While the Einstein equation becomes:
\beq
G_{\mu\nu}^{\rm dS} -{m_*^2\over 2} (h_{\mu\nu} - \gamma_{\mu\nu}h) -
{\cal {\hat O}}_{\mu\nu}\phi = -T_{\mu\nu}\,,
\label{Eq1}
\eeq
where the Einstein tensor on the dS space is defined in the usual way:
\beq
G_{\mu\nu}^{\rm dS} = {1\over 2} \left (\square h_{\mu\nu} -\nabla_\mu 
\nabla_\alpha h^\alpha_\nu - \nabla_\nu \nabla_\alpha h^\alpha_\mu + 
\nabla_\mu \nabla_\nu  h\right ) - 
{1\over 2} \gamma_{\mu\nu} \left (\square h -\nabla_\alpha \nabla_\beta 
h^{\alpha\beta}\right ) \nonumber \\  
 - H^2 \left ( h_{\mu\nu} + {1\over 2} 
\gamma_{\mu\nu} h \right )\,. 
\label{ET}
\eeq
Note that $\nabla^\mu G_{\mu\nu}^{\rm dS} =0$ and 
 $ \nabla ^\mu {\cal {\hat O}}_{\mu\nu}(\rm scalar)=0$, then, 
the Bianchi identities mentioned above follow from 
(\ref {Eq1})\footnote{Naively, there is no 
kinetic term 
for $\phi$, but it can be generated by
transformation (\ref{enhanced}) with $\rho =\alpha\phi $.
The additional term in the Lagrangian (\ref {eff0}) is 
\beq
\alpha 3H \phi (\Box+4H^2)\phi~. \nonumber
\eeq },  and $h=0$ from (\ref{Oh}). 
Therefore, $h_{\mu\nu}$ is transverse 
and traceless. Thus, out of the ten components of $h_{\mu\nu}$ only five
are unrestricted. These correspond to five degrees of freedom of a 
massive spin-2 state.  However, in addition there is a sixth degree of freedom
propagated by the $\phi$ field.  As we'll see below, this degree of freedom 
is ghost-like,  and moreover,  it does couple to the trace of the 
stress-tensor, making the theory inconsistent.

To calculate the field $h_{\mu\nu}$ produced by the source $T_{\mu\nu}$ 
it is useful to introduce a 
decomposition of the tensors in their transverse traceless 
(TT), pure trace (PT) and symmetric trace-free (ST) parts:
\beq
T_{\mu\nu} =T^{TT}_{\mu\nu} +{1\over 4}\gamma_{\mu\nu}T+{1\over 3}
P_{\mu\nu}{1\over Q}T\,,
\label{dec}
\eeq
and we use the notations
\beq
P_{\mu\nu}\equiv \nabla_\mu\nabla_\nu-{1\over 4}\gamma_{\mu\nu}\square,~~ 
{Q} \equiv -\square -4H^2,~~S\equiv -\square +4H^2.
\label{PQS}
\eeq
Noticing that 
\begin{equation}
P_{\mu\nu}(\square +8H^2)\varphi=\square  P_{\mu\nu}\varphi~,
\end{equation}
for any scalar $\varphi$, we get that $P_{\mu\nu}f(Q)\varphi=
f(S)P_{\mu\nu}\varphi$, for any smooth function $f$.

Next, let us introduce the Lichnerowicz operator  $\Delta_L$ 
which acts differently on the TT, PT and ST parts of the stress-tensor
\begin{eqnarray}\label{prop}
(\Delta_L -4H^2)T_{\mu\nu}^{TT}&=& S\, T_{\mu\nu}^{TT}~,\nonumber\\
(\Delta_L -4H^2)\gamma_{\mu\nu}\varphi&=&\gamma_{\mu\nu}Q
\varphi~,\\
(\Delta_L -4H^2)P_{\mu\nu}\varphi&=&P_{\mu\nu} Q \varphi~,\nonumber
\label{Lichn}
\end{eqnarray}
where  $\varphi$ is an arbitrary scalar.

Using the above relations, one can calculate the field. The result reads
\begin{eqnarray}
{1\over 2} h_{\mu\nu} =   {1\over\Delta_L -4H^2} T^{(1/3)}_{\mu\nu} 
 - {\gamma_{\mu\nu} \over 12} 
{T\over \square +4H^2} +
{1\over 3} \left ( {\nabla_\mu \nabla_\nu }  - {1 \over 4} 
\gamma_{\mu\nu} \square \right )
{T\over (\square +4H^2)^2}\,, 
\label{h0}
\end{eqnarray}
where we introduced the notation
\beq
T^{(1/3)}_{\mu\nu} \equiv T_{\mu\nu} - {1\over 3} 
\gamma_{\mu\nu} T , 
\label{T1/3}
\eeq
and all the operators act from the left to right.

It is also convenient to rewrite the expression 
(\ref {h0}) in the following form:
\beq
{1\over 2} h_{\mu\nu} =   {1\over\Delta_L -4H^2} T^{(1/3)}_{\mu\nu} 
-{1\over 3} \left ( {\nabla_\mu \nabla_\nu }  + {\gamma_{\mu\nu}
H^2} \right )
{T\over (\square +4H^2)^2}\,.
\label{h01}
\eeq

We can decipher the needed information 
about the spectrum from the expression (\ref {h01}):
The first term on the r.h.s. of (\ref {h01}) represents  
a contribution of the 
helicity-2 and helicity-0 components of the massive graviton.

In the second  term of the r.h.s. of  (\ref {h01}), 
there is a structure with covariant derivatives $\nabla_\mu \nabla_\nu$ 
that does not contribute to the gauge invariant physical 
amplitude,  ${\cal A} \equiv  \int  d^4 x \sqrt{\gamma}~
T^{\prime\mu\nu} h_{\mu\nu}$, because of the conservation 
of the stress-tensor $T_{\mu\nu}^{\prime}$.
Thus, in the linearized theory these terms carry no information 
(they will be important, though, in the nonlinear theory - see section 3).
However, there is a very last term in (\ref {h01}) which is proportional 
to $\gamma_{\mu\nu}$ and  contains a double pole $(\square +4H^2)^2$.
The latter can be decomposed into a sum of single poles
giving rise to a tachyon and a ghost-like state!

The goal of the next section will be  
to decouple these states entirely 
from the physical sector of the 
linearized theory.

\section{Decoupling the ghost}

To solve the problems of the previous section we will consider the 
Lagrangian of the tensor field $h_{\mu\nu}$ and scalar $\phi$ 
with the mixing and quadratic kinetic terms:
\beq\label{eff}
{\cal L}_{\rm eff}={\cal L}_*(h_{\mu\nu})-\phi 
{\cal {\hat O}}^{\mu\nu}
h_{\mu\nu}  +\phi {\cal {\hat K}}\phi+
h_{\mu\nu} T^{\mu\nu} +q \phi T,
\eeq
Here, ${\cal {\hat K}}$ is some operator
which can contain up to two derivatives as well as 
terms proportional to $H^2$ (will be specified below)
and $q$ is a constant determining the strength of the 
scalar coupling to the trace of the stress-tensor.

The Einstein  equation reads as follows:
\beq
G_{\mu\nu}^{\rm dS} -{m_*^2\over 2} (h_{\mu\nu} - \gamma_{\mu\nu}h) -
{\cal {\hat O}}_{\mu\nu}\phi = -T_{\mu\nu}\,,
\label{E1}
\eeq
while variation w.r.t.~$\phi$ gives
\beq
{\cal {\hat O}}^{\mu\nu}h_{\mu\nu}-2{\cal {\hat K}}\phi = qT.
\label{Phi1}
\eeq
The Bianchi identities give, as before 
$\nabla^\mu h_{\mu\nu} =\nabla_\nu h$, reducing the number of 
independent components of $h_{\mu\nu}$ from ten down to six.
Using the Bianchi identities in the equation of motion of 
$\phi$ (\ref {Phi1}) we find the equation that is linear in $h$, 
has no derivatives of $h$, and therefore, determines $h$ in terms of $\phi$. 
Finally the trace of the Einstein  equation determines $\phi$. 
The corresponding expression for $\phi$ and $h$ read:
\beq
\phi = {1\over 3Q}T\,, ~~~~~
\label{phiT}
h= -{1\over 3H^2} \left(q+{2\over 3}{{\cal \hat K}\over Q}\right)T~.  
\label{h22}
\eeq
Hence, there are six independent degrees of freedom, 
five in $h_{\mu\nu}$ and one in $\phi$. However, as we will
see below, for a special choice of the coefficient 
in (\ref {eff}) the sixth degree of freedom decouples from 
the physical sector in the linearized approximation. 
Alternatively, one could have excluded $\phi$ in favor of $h$ suing 
(\ref {Phi1}) and the Bianchi identities; then, 
there would  remain  six unconstrained degrees of freedom in 
$h_{\mu\nu}$.

The physical metric that couples to the source, 
and determines the physical amplitude is:
\beq
h^{\rm pys}_{\mu\nu} = h_{\mu\nu} + \gamma_{\mu\nu} q \phi =
h_{\mu\nu}  - \gamma_{\mu\nu} q {T\over 3 (\square +4H^2)}\,.
\label{physh}
\eeq

After some algebra, which  uses  the properties of 
the Lichnerowicz operator (\ref {Lichn}),  
one can find the following expression for the physical field
\beq
{1\over 2} h_{\mu\nu}^{\rm phys}= {1\over\Delta_L -4H^2} T^{(1/3)}_{\mu\nu} 
+ \gamma_{\mu\nu}\left (q -a\right )
{T\over 3Q}+
{\nabla_\mu \nabla_\nu \over 3H^2} \left ({q\over 2} +a\right )
{T\over Q}\,,
\label{hphysfin}
\eeq
where we have used an explicit form for ${\cal \hat K}$, namely:
\beq
{\cal {\hat K}} = 3H^2+3aQ\,.  
\label{K}
\eeq
The first term in (\ref{K})
is necessary to cancel the double pole!

In (\ref {hphysfin}) there is still freedom in choosing $a$. This should 
be used to remove the  tachyonic pole in the second term in 
(\ref {hphysfin}). Hence, we can set
$a=q$. Then the expression for the physical metric reads:
\beq
{1\over 2} h_{\mu\nu}^{\rm phys}= {1\over\Delta_L -4H^2} T^{(1/3)}_{\mu\nu} 
-{q\over 2} {\nabla_\mu \nabla_\nu \over H^2} {T\over \square +4H^2}\,.
\label{hphy2}
\eeq
We achieved our goal -- there are no ghost or tachyons in  the part 
of (\ref {hphy2}) that couples to a conserved stress-tensor; 
the last term in (\ref {hphy2}) does not contribute to 
physical amplitudes in the linearized approximation. That term  
would,  however, contribute to Feynman diagrams with 
nonlinear self-interactions of gravitons as long as $q\neq 0$.
It would give rise to strong  coupling behavior in a 
full non-linear theory. Then, it should  be possible to reconcile the 
predictions of such a theory with the observations
\cite {Arkady,DDGV} (for a brief 
review, see \cite {GIrev}). However, for $q\neq 0$ there might 
be a danger that the sixth degree of freedom would contribute
to the physical sector in the full nonlinear theory. 
This, however, may be possible to avoid if $\phi$ corresponds to a 
Nambu-Goldstone boson of a nonlinearly realized symmetry,
similar to the brane bending mode in brane induced 
gravity. 

Alternatively, for $q=0$  one would not expect to have 
strongly coupled  behavior in the nonlinear theory.
This would be similar to ``softly massive'' gravity  
\cite {GShifman} (see also \cite {Massimo}).
Then, one would have to overcome the vDVZ discontinuity
\cite {vDVZ}. Some new ideas are needed for this.

It is also possible that there exist  nonlinear theories 
that give rise to  selfaccelerated backgrounds, which do  not 
necessarily respect the relation (\ref {mstar0}). Some of these
might  be possible to construct by pursuing 
further the proposal of Refs. \cite {GGn}.

\vspace{0.2in}

The work  of GG was supported by  NASA grant NNGG05GH34G. AI was 
supported by DOE grant DE-FG03-91ER40674.


\end{document}